\documentclass[a4paper,dvips,11pt]{article}

\usepackage{cite}
\usepackage{epsf}
\usepackage{multirow} 
\usepackage{bm,bbm} 
\usepackage[pdftex]{graphicx}
\usepackage{tabularx}
\usepackage{latexsym}
\usepackage{amsmath,amssymb,exscale}
\usepackage{array,multicol}
\numberwithin{equation}{section}

\def\draftlabel#1{{\@bsphack\if@filesw {\let\thepage\relax
   \xdef\@gtempa{\write\@auxout{\string
      \newlabel{#1}{{\@currentlabel}{\thepage}}}}}\@gtempa
   \if@nobreak \ifvmode\nobreak\fi\fi\fi\@esphack}
        \gdef\@eqnlabel{#1}}
\def\@eqnlabel{}
\def\@vacuum{}
\def\draftmarginnote#1{\marginpar{\raggedright\scriptsize\tt#1}}
\def\draft{\oddsidemargin -.5truein
        \def\@oddfoot{\sl preliminary draft \hfil
        \rm\thepage\hfil\sl\today\quad\militarytime}
        \let\@evenfoot\@oddfoot \overfullrule 3pt
        \let\label=\draftlabel
        \let\marginnote=\draftmarginnote
   \def\@eqnnum{(\theequation)\rlap{\kern\marginparsep\tt\@eqnlabel}%
\global\let\@eqnlabel\@vacuum}  }

\newcommand{\PRL}[3]{\emph{ Phys.~Rev.~Lett.} \textbf{#1} (#2) #3}

\newcommand{\PR}[3]{\emph{ Phys.~Rep.} \textbf{#1} (#2) #3}

\def\ov{\overline}

\def\dalemb#1#2{{\vbox{\hrule height .#2pt
         \hbox{\vrule width.#2pt height#1pt \kern#1pt
                 \vrule width.#2pt}
         \hrule height.#2pt}}}

\def\half{{\textstyle{1\over2}}}
\let\a=\alpha    
    
    \let\p=\pi 
\let\s=\sigma     

      \let\G=\Gamma  
  \let\S=\Sigma   
\let\F=\Phi

 \def\bd{\begin{document}} \def\ed{\end{document}}
\def\ds{\documentstyle} \let\fr=\frac \let\bl=\bigl \let\br=\bigr
\let\Br=\Bigr \let\Bl=\Bigl
\let\bm=\bibitem
\let\na=\nabla
\let\pa=\partial
\let\ov=\overline
\def\ie{{\it i.e.\ }}
\def\tr{{\mbox{\rm tr}}}
\newcommand{\be}{\begin{equation}}
\newcommand{\ee}{\end{equation}}
\newcommand{\beba}{\begin{equation}\begin{array}{lcl}}
\newcommand{\eaee}{\end{array}\end{equation}}
\newcommand{\bea}{\begin{eqnarray}}
\newcommand{\eea}{\end{eqnarray}}
\newcommand{\ba}{\begin{array}}
\newcommand{\ea}{\end{array}}
\newcommand{\td}{\tilde}
\newcommand{\norsl}{\normalsize\sl}
\newcommand{\ns}{\normalsize}
\newcommand{\refs}[1]{(\ref{#1})}
\def\simlt{\mathrel{\lower2.5pt\vbox{\lineskip=0pt\baselineskip=0pt
            \hbox{$<$}\hbox{$\sim$}}}}
\def\simgt{\mathrel{\lower2.5pt\vbox{\lineskip=0pt\baselineskip=0pt
            \hbox{$>$}\hbox{$\sim$}}}}
\def\A{{\cal A}}
\def\a{{\mathcal a}}
\def\V{{\cal V}}
\def\F{{\cal F}}
\def\p{{\mathcal \phi}}
\def\L{{\mathcal L}}
\def\M{{\mathcal M}}
\def\bD{{\ov {\rm D}}}
\def\bO{{\ov {\rm O}}}
\def\bOp{{\ov {\rm O'}}}
\def\O{{ {\rm O}}}

\newcommand{\nsect}{\setcounter{equation}{0}
\def\theequation{\thesection.\arabic{equation}}\section}
\newcommand{\nappend}{\setcounter{equation}{0}
\def\theequation{\rm{A}.\arabic{equation}}\section*}
\newcommand{\appendixA}{\setcounter{equation}{0}
\def\theequation{\rm{A}.\arabic{equation}}\section*}
\newcommand{\appendixB}{\setcounter{equation}{0}
\def\theequation{\rm{B}.\arabic{equation}}\section*}
\newcommand{\appendixC}{\setcounter{equation}{0}
\def\theequation{\rm{C}.\arabic{equation}}\section*}
\newcommand{\appendixD}{\setcounter{equation}{0}
\def\theequation{\rm{D}.\arabic{equation}}\section*}
\newcommand{\appendixE}{\setcounter{equation}{0}
\def\theequation{\rm{E}.\arabic{equation}}\section*}
\newcommand{\appendixF}{\setcounter{equation}{0}
\def\theequation{\rm{F}.\arabic{equation}}\section*}
\newcommand{\appendixG}{\setcounter{equation}{0}
\def\theequation{\rm{G}.\arabic{equation}}\section*}
\def\baselinestretch{1.5}
\catcode`\@=11
\def\marginnote#1{}
\newcount\hour
\newcount\minute
\newtoks\amorpm
\hour=\time\divide\hour by60
\minute=\time{\multiply\hour by60 \global\advance\minute by-\hour}
\edef\standardtime{{\ifnum\hour<12 \global\amorpm={am}%
        \else\global\amorpm={pm}\advance\hour by-12 \fi
        \ifnum\hour=0 \hour=12 \fi
        \number\hour:\ifnum\minute<10 0\fi\number\minute\the\amorpm}}
\edef\militarytime{\number\hour:\ifnum\minute<10 0\fi\number\minute}
\def\draftlabel#1{{\@bsphack\if@filesw {\let\thepage\relax
   \xdef\@gtempa{\write\@auxout{\string
      \newlabel{#1}{{\@currentlabel}{\thepage}}}}}\@gtempa
   \if@nobreak \ifvmode\nobreak\fi\fi\fi\@esphack}
        \gdef\@eqnlabel{#1}}
\def\@eqnlabel{}
\def\@vacuum{}
\def\draftmarginnote#1{\marginpar{\raggedright\scriptsize\tt#1}}
\def\draft{\oddsidemargin -.5truein
        \def\@oddfoot{\sl preliminary draft \hfil
        \rm\thepage\hfil\sl\today\quad\militarytime}
        \let\@evenfoot\@oddfoot \overfullrule 3pt
        \let\label=\draftlabel
        \let\marginnote=\draftmarginnote
   \def\@eqnnum{(\theequation)\rlap{\kern\marginparsep\tt\@eqnlabel}%
\global\let\@eqnlabel\@vacuum}  }

\def\preprint{\twocolumn\sloppy\flushbottom\parindent 1em
        \leftmargini 2em\leftmarginv .5em\leftmarginvi .5em
        \oddsidemargin -.5in    \evensidemargin -.5in
        \columnsep 15mm \footheight 0pt
        \textwidth 250mmin      \topmargin  -.4in
        \headheight 12pt \topskip .4in
        \textheight 175mm
        \footskip 0pt
        \def\@oddhead{\thepage\hfil\addtocounter{page}{1}\thepage}
        \let\@evenhead\@oddhead \def\@oddfoot{} \def\@evenfoot{} }

\def\titlepage{\@restonecolfalse\if@twocolumn\@restonecoltrue\onecolumn
     \else \newpage \fi \thispagestyle{empty}\c@page\z@ 
        \def\thefootnote{\fnsymbol{footnote}} }

\def\endtitlepage{\if@restonecol\twocolumn \else  \fi
        \def\thefootnote{\arabic{footnote}}
        \setcounter{footnote}{0}}  

\catcode`@=12
\relax
\def\abs#1{\left| #1\right|}
\def\bC{\mathop{\bf C}}
\def\bea{\begin{array}}
\def\bem{\begin{displaymath}}
\def\beq{\begin{equation}}
\def\bea{\begin{eqnarray}}
\def\bR{\mathop{\bf R}}
\def\bra#1{\left\langle #1\right|}
\def\eea{\end{array}}
\def\eem{\end{displaymath}}
\def\eeq{\end{equation}}
\def\eea{\end{eqnarray}}
\def\eq{\beq\eeq}                          
\def\eqr#1{\beq\label#1\eeq}               
\def\half{\frac{1}{2}}
\def\Im{\mathop{\rm Im}}
\def\ket#1{\left| #1\right\rangle}
\def\sket#1{| #1 >}
\def\lie{\hbox{\it \$}}                          
\def\lineint{\oint \frac{d z}{2 \pi i}} 
\def\modsq#1{| #1 |^2}
\def\NP#1#2#3{Nucl. Phys. \underline{#1} (19#2) #3}
\def\ov{\overline}
\def\partder#1#2{{\partial #1\over\partial #2}}
\def\PL#1#2#3{Phys. Lett. \underline{#1} (19#2) #3}
\def\PR#1#2#3{Phys. Rev. \underline{#1} (19#2) #3}
\def\PRL#1#2#3{Phys. Rev. Lett. \underline{#1} (19#2) #3}
\def\Re{\mathop{\rm Re}}
\def\secder#1#2#3{{\partial^2 #1\over\partial #2 \partial #3}}
\def\s2w{\sin^2 \theta_W}
\def\Tr{\mathop{\rm Tr}}
\def\und{\underline}
\def\VEV#1{\left\langle #1\right\rangle} \let\vev\VEV
\def\mbf#1{\hbox{\boldmath $#1$}}
\relax
\def\dalpha{{\dot\alpha}}
\def\dbeta{{\dot\beta}}
\def\drho{{\dot\rho}}
\def\dsigma{{\dot\sigma}}
\def\crbig{\\\noalign{\vspace {3mm}}}
\def\bigint{{\displaystyle\int}}
\def\S{\Sigma}
\def\G{\Gamma}
\def\L{{\cal L}}
\def\SG{S_{\Gamma}}
\def\Fint{{\bigint d^2\theta\,}}
\def\Fbarint{{\bigint d^2\ov\theta\,}}
\def\Dint{{\bigint d^2\theta d^2\ov\theta\,}}

\title{ \vspace*{-0.8cm}
\begin{flushright}
\normalsize{CPTH-RR-071-0707\\}
\end{flushright}
\vspace{1cm}
\bf{Toward metastable string vacua\\ from magnetized branes} \vspace*{-0.3cm}}

\date{}
\begin{document}
\author{\bf\large{ T. Maillard$^{1}$\footnote{Tristan.Maillard@cpht.polytechnique.fr}}\\  
\\[-3mm]
\emph{\normalsize $^1$ CPHT, UMR du CNRS 7644 , Ecole Polytechnique, 91128 Palaiseau, France}\\
}

\date{}

\maketitle
\thispagestyle{empty}

\begin{abstract}

The scalar potential of recombination fields of magnetized branes in Type IIB orientifold compactifications  is analyzed in the absence of any closed string fluxes. Considering its  perturbative $F$ and $D$-term contributions in a quadratic approximation, we present the conditions for which its minima are supersymmetric.  We show that for reasonable conditions on the spectrum, both metric moduli and recombination fields can be stabilized. We then provide explicit examples of compact manifolds where a Minkowski vacuum  is realized in a Higgs phase. The vacuum energy is zero and some charged scalars acquire a vev. We then address the question of supersymmetry breaking.  The scalar potential for recombination fields is analyzed when supersymmetry is broken by $F$ and $D$-term. We show that locally stable vacua can exist at the classical level. These are formed by a {\it hidden } supersymmetric sector that fixes  metric moduli and recombination fields and a {\it visible}  sector  where supersymmetry is spontaneously broken.

\end{abstract}

\date


\maketitle
 \vspace*{-0.8cm}
\hrulefill
\newpage
\section{Introduction}

String theory is known to possess a large number of vacua which reproduce the basic properties of the standard model of particle physics \cite{Douglas:2006es}. In particular in Type IIB string theory, gauge group, chirality and family replication, can be described by magnetized branes \cite{Review,openAS}. However, these vacua generally depend on continuous parameters that correspond to vacuum expectation values (vev) of so-called moduli fields, characterizing for instance the size and shape of the compactification manifold.  These are perturbatively flat directions of the scalar potential. However,  precision measurements of the principle of equivalence strongly constraint the existence of light scalars \cite{Will:2001mx}.  Moreover, the strengths of the interactions and the mass spectrum in the low energy effective action of string theories depend on the vev of these moduli, leading to a loss of predictivity.  It is therefore of great interest to understand the mechanism of stabilization of the string vacuum.

Similarly, flat compactifications of type II orientifolds contain a number of massless scalar states in the open string sector. They describe either the geometrical deformations of single branes, Wilson lines or the recombination  of brane configurations. The lifting of a potential and the existence of non-trivial minima for these fields are of great importance, since  linked with the Higgs sector of the standard model. For geometrical moduli,  stabilization mechanisms have been proposed in \cite{Gomis:2005wc,Lust:2005bd,Angelantonj:2003zx}. Here, we focus on perturbative stabilization of the {\it recombination} fields. 

In Type IIB orientifold compactifications, these are associated with the lowest excitations of open strings stretched between two distinct magnetized branes\cite{Review,openAS}. In the bifundamental  or (anti)-symmetric representation of the gauge group, they are generically massive or tachyonic, indicating an unstable vacuum. However, for suitable choices of the magnetic fluxes and volume moduli, the lightest mode becomes massless and supersymmetry is restored\cite{Angelantonj:2000hi,Blumenhagen:2000wh}. We show here that magnetized branes perturbatively generate a potential for the recombination fields as well as the  metric moduli. Their masses are strictly positive and their vev  are stabilized at non-zero values.

To that end,  combined effects of the $F$ and $D$-terms in the scalar potential are considered, similarily as  \cite{Cremades:2007ig}. Yet,  only perturbative contributions to the potential in absence of closed string fluxes are considered here.  It is indeed  well known that trilinear terms appear in the superpotential of the magnetized branes' effective action \cite{Cremades:2004wa}. Aside from Yukawa couplings, this induces quartic couplings in the scalar potential that depend exclusively on the complex structure. Similarly, for each anomalous abelian factor of the gauge group, there exists a Fayet-Iliopoulos (FI) parameter in the $D$-term. \cite{Dine:1987xk,Atick:1987gy}.  Thanks to its K\"ahler dependance, it has already been shown that K\"ahler moduli can be perturbatively stabilized in absence of any recombination fields \cite{BLT}. However, pure $D$-term stabilization leads to flat directions which correspond to combinations of the open string and K\"ahler moduli \cite{Cvetic:2001nr}.

These flat directions can also be lifted in a supersymmetric  vacuum where all $F$ and $D$-terms vanish. Under minimal assumptions on the flavor spectrum, most of the recombination fields are stabilized at zero vev by $F$-flatness conditions. The remaining ones are in turn stabilized by $D$-flatness conditions at non-zero values. All of them then acquire a positive mass.  Since some scalars have non-zero vevs, the gauge group is spontaneously broken. Geometrically, this may be described by  the recombination of different stacks of  branes that correspond, in the T-dual picture, to instanton transitions of branes at angles \cite{Cvetic:2001nr}. Moreover, when the number of supersymmetry constraints is bigger than the number of  recombination fields, the consistency of the different $D$-flatness conditions also restricts the K\"ahler moduli \cite{BLT,AM, AKM2}. Finally, $F$-flatness conditions arising from geometrical moduli restrict the complex structure. 

 Note that no three-form fluxes are needed here. They would induce soft terms in the superpotential that may lead, together with $D$-terms,  to perturbative K\"ahler moduli stabilization\cite{GarciadelMoral:2005js}, but are  strongly constrained by Freed-Witten anomaly \cite{Freed:1999vc}. Therefore there exists, in a Higgs phase, configurations of branes that satisfy all RR tadpole conditions and stabilize most of the recombination and metric moduli in a supersymmetric vacuum. The presence of magnetized branes then not only determines   the gauge symmetry breaking pattern, but also the size and shape of the underlying geometry, confirming  the mechanism proposed by \cite{AM, AKM2}.

Beyond this global minimum, there may exist local minima. Indeed, other configurations of magnetized branes  exist for which supersymmetry cannot be globally restored. The system of supersymmetric conditions arising from distinct stacks of branes is overconstrained. In this case, there are no supersymmetric vacua. Yet, we will show that there are points in the closed string moduli space where classically stable  non-supersymmetric vacua appear. These local minima are made of two sectors that we will designate as {\it visible} and {\it hidden}. In the first one, all $F$ and $D$-terms vanish and supersymmetry is restored. Similarly to  the previous case, recombination fields and metric moduli are stabilized but the supersymmetric sector does not satisfy all tadpole conditions by itself.  In order to form a consistent vacuum, a second  sector is then added. For some flux quanta, supplementary  $F$ and $D$-flatness conditions are incompatible with the initial one.  Supersymmetry is then broken. Most of the supersymmetry breaking configurations of magnetized branes have tachyonic directions in the open string sector. However, there are  windows in the closed string moduli space where all recombination fields of the {\it hidden} sector acquire a positive mass. Classically, the configuration is then stable. 

These kinds of models belong to the class of non-supersymmetric  tachyon-free constructions  that are  possible in a small window of the K\"ahler moduli space. However,   even if all RR-tadpoles are cancelled, the supersymmetry breaking implies a non-vanishing tadpole for the dilaton. Indeed  the value of the scalar potential at these local minima are  positive. Since it comes from  a disk amplitude, it corresponds to a runaway potential for the dilaton. This problem may be turned into an advantage. Combined with effects on different orders in the string amplitude, a dilaton stabilization and  a potential uplift may be achieved  \cite{Cremades:2007ig,ADM}.  Yet, this goes beyond the scope of this work and will be discussed in \cite{ADM}.

The article is organized in the following way: In section \ref{basic}, the basic idea of moduli stabilization by $F$- and $D$-terms  in field theory  is introduced. It is shown how FI-parameters are constrained in a supersymmetric vacuum, but also under what conditions a non-supersymmetric vacuum may exist.  In section \ref{string}, we  then  present how this idea may be implemented to magnetized branes in Type IIB orientifold compactifications. Both the spectrum and the dependance of the trilinear coupling and FI-parameters on the closed string moduli and flux quanta, and the constraints on the flux coming from the tadpole conditions are reviewed. The structure of the scalar potential is presented in section \ref{ScPot} in the string compactification as well as its minima and mass terms. Finally, explicit examples of supersymmetric  are given in section \ref{sec:susy}.

\section{Basic Setup}\label{basic}

In this section, basic ingredients for stabilization of open string moduli are presented. We aim at showing that the typical scalar potential that appears in the effective action of magnetized branes may achieve a full stabilization of open string moduli by the combined effects of $F$ and $D$-terms. Trilinear couplings in the superpotential and non-vanishing FI-terms in the $D$-terms give rise to positive mass for the open string moduli. Depending on the values of the FI-terms and trilinear coupling constants,  both supersymmetric and non-supersymmetric vacua can be achieved in a Higgs phase. To this end, the gauge symmetry is chosen to be a product of unitary groups and the matter content is either in bifundamental or antisymmetric representations. For both cases, we show how they acquire a non-vanishing vev and a positive mass. 

Note that the gauge groups considered here are  anomalous in general. However, since the mechanism presented here will be embedded in a string construction, the corresponding gauge bosons will acquire a mass via the Green-Schwarz (GS) mechanism\cite{Review}.

\subsection{Supersymmetric vacua}\label{abelian}

Let us consider first  the easiest theory that possesses a supersymmetric vacuum in which all charged fields acquire a positive mass. It is formed by the product of three abelian gauge groups  $G_1=U(1)\times U(1)\times U(1)$ with chiral  fields in bifundamental representations as given in Table \ref{triangle}.
\begin{table} 
\begin{center}
\begin{tabular}{c|ccc}
 & $\phi_{12}$  &$\phi_{31}$  & $\phi_{23}$ \cr
\hline

$U(1)$ & $1$  & $-1$ &$0$ \cr

$U(1)$ & $-1$  & $0$ &$1$  \cr

$U(1)$& $0$  & $1$ &$-1$\cr
\end{tabular}
\vskip .4cm
\end{center}
\caption{Triangle of chiral fields in bifundamental representations of abelian gauge groups.}
\label{triangle}
\end{table}
The most general tree-level superpotential consistent with the symmetries $G$ is given by the trilinear term 
\be
W = W_{123}\; \phi_{12}\phi_{23}\phi_{31} \, ,
\label{trilinear}
\ee
while the $D$-term for each $U(1)$ factor reads
\be
D_a = \left(\sum_{i=1}^3q_i^a |\phi_i|^2 + \xi_a\right) \quad \quad  , \quad \quad a=1,2,3 \; , 
\label{D1}
\ee
where $\xi_a$ are the FI-parameters and $q_i^a=\pm 1$  the charges of the fields $\phi_i$ in respect to the $a$-th abelian factor. 
Assuming a canonical K\"ahler potential, $F$ and $D$- flatness conditions $\partial_i W=W=0$ and  $D_a=0$, $\forall a, i=1,2,3$ can only be satisfied if the FI-parameters satisfy 
\be
\xi_3=0 \quad \quad {\rm and}\quad \quad \xi_1=-\xi_2<0 \, .
\label{FIsusy}
\ee
In this case, the charged fields $\phi_i$ have a minimum at 
\be
\langle\phi_{31}\rangle=\langle\phi_{23}\rangle=0\quad \quad {\rm and}\quad \quad \langle|\phi_{12}|^2\rangle :=v^2= -\xi_1 \, .
\ee 
where their masses are determined by the trilinear coupling $W_{123}$ and the FI-parameter $\xi_1$ as
\be
M_{\phi_{31}}^2 \sim |W_{123}|^2|\xi_1| \quad , \quad M_{\phi_{23}}^2 \sim |W_{123}|^2|\xi_1|  \quad , \quad M_{\phi_{12}}^2 \sim |\xi_1|    \, .
\ee
In this toy model, a supersymmetric vacuum exists  only if the FI-parameters are in the domain (\ref{FIsusy}). The full moduli space is then restricted to a point at which one  of the three scalars obtains a non-vanishing vev and a mass that are proportional to the non-vanishing FI-parameters. The masses of the two  remaining fields are also proportional to the trilinear coupling.  Indeed, for magnetized branes, the FI-parameters  depend on the volume moduli. The constraints (\ref{FIsusy}) can therefore be interpreted as conditions on the volume moduli. The combined effects of $F$ and $D$- terms then lead to the stabilization not only of  the twisted open string moduli  but also of the K\"ahler moduli. 

Note that a hierarchy in the scalar  mass spectrum may exist. Indeed,  contrarily to the Higgs fields, the scalar fields whose vev is stabilized at zero values depend on the trilinear coupling $h^2$. For  instance, very small $h^2$ would imply Higgs masses much larger than the other scalar masses.

The first example was restricted to three abelian gauge symmetry with a minimal number of matter fields in bifundamental representations $(\mathbf{N}_a,\bar{\mathbf{N}}_b)$, which are typically  of compactifications of magnetized branes in Type IIB compactifications. In orientifold compactifications however,   another class of interesting chiral fields arises. These are either  in bifundamental $(\mathbf{N}_a,\mathbf{N}_b)$  or (anti)-symmetric ($A_a$) $S_a$ representations of the gauge group.  The stabilization mechanism remains similar, but the restrictions on the FI-parameters change drastically.  Let us postpone the discussion on antisymmetric representations and analyze the square of abelian gauge groups $G=U(1)\times U(1)\times U(1)\times U(1)$ with chiral fields in all possible bifundamental representations as given in Table \ref{table:4abelian}. The most general $D$-term remains identical as in eq (\ref{D1}), but  the superpotential  reads
\bea
W^{(4)} &=&  \sum_{i,k}\, W^{\; ik}_1 \; \phi_{12}\phi^k_{23}\phi^i_{31} + \sum_{j,m}\, W^{\; jm}_{2}\;\phi_{12}\phi^m_{24}\phi^j_{41} 
\nonumber
\\
&+&  \sum_{i,k}\, W^{\; i^\star k^\star}_{1^\star} \phi_{12}\phi^{i^\star}_{23^\star}\phi^{k^\star}_{3^\star 1} + \sum_{j,m}\, W^{\; j^\star m^\star}_{2^\star}\phi_{12}\phi_{24^\star}^{m^\star}\phi_{4^\star 1}^{j^\star} \, ,
\label{supot4}
\eea
where the indices $i,j,k,m$ and $i^\star,j^\star,k^\star,m^\star$ denote the flavor of the different matter fields. For instance, $i =1,\dots,I_{31} $ and $i^\star = 1,\dots,I_{3^\star 1}$, where $I_{ab}$ and $I_{a^\star b}$ are defined to be the number of flavors in $(\mathbf{N}_a,\bar{\mathbf{ N}}_b)$ and $(\mathbf{N}_a,\mathbf{N}_b)$. 
\begin{table}
\begin{center}
\begin{tabular}{c|cccccc|cccccc}
 & $\phi_{12}$ & $\phi^i_{31}$ &$\phi^j_{41}$ & $\phi^k_{23}$&$\phi^m_{24}$& $\phi_{34}$& $\phi_{12^\star}$ & $\phi^{i^\star}_{3^\star1}$ &$\phi^{j^\star}_{4^\star1}$ & $\phi^{k^\star}_{23^\star}$&$\phi_{24^\star}^{m^\star}$& $\phi_{34^\star}$\\
\hline
 $U(1)$&1& -1  & -1 &0 &0&0&1& -1  & -1 &0 &0&0\\
$U(1)$&-1& 0  & 0&1&1&0&1& 0  & 0&1&1&0 \\
$U(1)$&0& 1  & 0 &-1&0&1&0& -1  & 0 &1&0&1\\
$U(1)$&0& 0  & 1 &0&-1&-1&0& 0  & -1 &0&1&1\\
\end{tabular}
\vskip .4cm
\end{center}
\caption{All possible bifundamental matter $\phi_{ab}$ and $\phi_{ab^\star}$ in $(\mathbf{N}_a,\bar{\mathbf{N}}_b)$ and  $(\mathbf{N}_a,\mathbf{N}_b)$ representations of four abelian gauge factors. The upper indices $i,j,k,m$ and $i^\star,j^\star,k^\star,m^\star$ are flavor indices.}
\label{table:4abelian}
\end{table}
\\
In this theory, there exists a supersymmetric vacuum where four charged fields remain unconstrained by the $F$-flatness conditions. Take for instance $\phi_{12}$, $\phi_{12^\star}$, $\phi_{34}$ and  $\phi_{34^\star}$. These fields acquire a positive mass and  a non-vanishing vev from the four $D$-terms that reads
\be
\begin{array}{c}
\langle |\phi_{12^\star}|^2\rangle \sim \xi_2-\xi_1 \quad {\rm and} \quad \langle |\phi_{12}|^2\rangle \sim -\xi_1-\xi_2
\\
\langle |\phi_{34}|^2\rangle \sim \xi_4-\xi_3 \quad {\rm and} \quad \langle |\phi_{3 4^\star}|^2\rangle \sim -\xi_4-\xi_3 \, , 
\end{array}
\label{orient:vev}
\ee 
while  all other fields are stabilized at zero vev from $F$-terms with masses   given by
\be
\begin{array}{l}
M_{ii\prime}^2 \sim \langle |\phi_{12}|^2  \rangle \left( W_1 W_1^\dagger \right)_{ii\prime}\; , \; M_{kk\prime}^2 \sim \langle |\phi_{12}|^2  \rangle \left( W_1^T W_1^\star \right)_{kk\prime}
\\
M_{jj\prime}^2 \sim \langle |\phi_{12}|^2  \rangle \left( W_2W_2^\dagger \right)_{jj\prime} \; , \;  M_{mm\prime}^2 \sim \langle |\phi_{12}|^2  \rangle \left( W_2^T W_2^\star \right)_{mm\prime}
\end{array}
\ee
and similarily for the other fields in $(\mathbf{N}_a,\mathbf{N}_b)$ representations. The constrains on FI-parameters $\xi_a$ are different than  in eq (\ref{FIsusy}). Here, the positivity of the vevs of eq (\ref{orient:vev}) demands that
\be
\xi_{2i} \pm \xi_{2i-1} <0 \quad {\rm for } \quad i=1,2
\label{Orient:FIcondGenOrient}
\ee
A full stabilization  of the open string moduli is then possible under a few assumptions. On one hand, there  must only be a single  flavor in $(\mathbf{N}_{2i-1},\bar{\mathbf{N}}_{2i})$ and $(\mathbf{N}_{2i-1},\bar{\mathbf{N}}_{2i})$ representations, for $i=1,2$. Indeed, their stabilization is achieved by $D$-terms. The number of stabilized fields can therefore be at most the number of abelian factors.  Additional flavors in these representations would remain flat directions of the scalar potential. The number of flavors in other representations can however be arbitrary, since a mass for each of them will be lifted by $F$-terms. On the other hand, the FI-parameters of the different $U(1)_a$-factors must be in domain (\ref{Orient:FIcondGenOrient}).  This has an importance consequence on the K\"ahler moduli stabilization that will be discussed in section \ref{ScPotSUSY}

Let us now present the stabilization of antisymmetric scalars. They are present in the spectrum only if the gauge group is $SU(N)$, for $N\geq 2$. Let us therefore consider the gauge symmetry $G=U(1)\times [SU(2)\times U(1)]$ with chiral fields $\phi_{12}$, $\phi_{2^\star 1}$ and $\phi_{2\, 2^\star}$ in $(1,\bar{\mathbf{2}})$, $(-1,\bar{\mathbf{2}})$ and $A_2$ representations respectively, the superpotential reads 
\be
W = h\; \phi_{12}\phi_{2^\star 1}\phi_{22^\star} \, , 
\label{supotorient}
\ee
where the terms have been correctly antisymmetrized. By $F$-flatness conditions, the fields $\phi_{12}$ and $\phi_{2^\star 1}$ have zero vev and consequently, the $D$-flatness conditions constrain the antisymmetric field $\phi_{22^\star}$ to have a non-vanishing vev and the FI-parameters to be in the domain where
\be
\xi_1 =0 \quad {\rm and} \quad \langle |\phi_{22^\star } |^2\rangle = - {\xi_{2}\over 2} >0 \, .
\label{antisymFIcond}
\ee

\subsection{Non-supersymmetric vacua}\label{flavors}

Let us now analyze the possibility that the typical  perturbative potential of magnetized branes has locally stable minima where supersymmetry is broken by $F$ and $D$ terms. In the parameter space where FI-parameters $\xi_a$ do not sit in domains (\ref{orient:vev}) or  (\ref{antisymFIcond}), no supersymmetric vacua exist.  Even if one may expect that non-supersymmetric configurations are not stable, we will see here that there are points of the parameter space where the vacua are non-supersymmetric and where all charged fields classically  acquire a positive mass. This depends on the parameter space spanned by the trilinear couplings $W^{ijk}$ and FI-parameters $\xi_a$.

Let us first  restrict  to the case of three abelian symmetries with $(\mathbf{N}_a,\bar{\mathbf{N}}_b)$  bifundamental matter fields. We analyze the vacua where $\langle W\rangle =0$. In this case, the scalar potential 
\bea
V(\phi_i) &=&h^2 \, |\phi_{12}|^2|\phi_{31}|^2 + 
\left(|\phi_{12}|^2-|\phi_{31}|^2 +\xi_1\right)^2+\left(-|\phi_{12}|^2+\xi_2\right)^2 +\left(|\phi_{31}|^2 +\xi_3\right)^2
\label{flavors:ScPot}
\eea
for the superpotential (\ref{trilinear})  and spectrum given in Table \ref{triangle}  possesses a minimum when the FI-parameters and trilinear coupling are in a domain where
\be
\left(h^2+2\right) (\xi_3-\xi_1) >4(\xi_1-\xi_2) \quad\quad  , \quad \quad (\xi_2-\xi_3) > 0 \quad {\rm and } \quad h^2<6 \, .
\label{NSusyCond}
\ee
The coupling constant $h^2$ is defined as  $h^2 = |W_{123}|^2$. Note that to make it simple, it is assumed that all gauge couplings are the same,  $g_a\equiv g =1$. The charged fields are fixed at the values $\langle\phi_{23}\rangle=0$ and 
 \be
 \begin{array}{c}
{1\over 2}(h^2+2)(6-h^2)\langle|\phi_{12}|^2\rangle=h^2 (\xi_3-\xi_1) + 4(\xi_2-\xi_1)-2(\xi_3-\xi_1)\, .
 \\
{1\over 2} (h^2+2)(6-h^2)\langle|\phi_{31}|^2\rangle=h^2 (\xi_1-\xi_2) + 4(\xi_1-\xi_3)-2(\xi_1-\xi_2)
 \end{array}
 \ee
 At this point of the moduli space, the eigenvalues of the mass matrix
 \be
  \begin{array}{c}
M_{\phi_{23}}^2 \sim (\langle D_2 \rangle- \langle  D_3 \rangle)>0 
\\
\\
 M_{12,31}^2 \sim \left(\langle|\phi_1|^2\rangle+\langle|\phi_2|^2\rangle\right)\left(1 \pm \sqrt{1-\epsilon} \right)>0  \quad \quad {\rm where}\quad \quad
 \\
 \\
 \epsilon = {\langle|\phi_1|^2\rangle\langle|\phi_2|^2\rangle\over \langle|\phi_1|^2\rangle+\langle|\phi_2|^2\rangle}(2+h^2)(6-h^2)<1 \, .
 \end{array}
 \ee
 are all positive. The F-term of the field $\phi_{23}$  does not vanish, $F_{\phi_{23}}\neq 0$ while the $D$-term of all three abelian factors is also non-zero. Supersymmetry is then spontaneously broken.  It is therefore interesting to note that this sector is classically  stable even if  supersymmetry is spontaneously broken. Yet,  in the domain where  the $\xi_a$'s and $h^2$ satisfy  eq. (\ref{NSusyCond}), the scalar potential  receives radiative corrections. The corrected potential is however not expected to develop tachyonic directions.
 
	In addition to scalars in $({\mathbf{N}}_a, \bar{\mathbf{N}}_b)$ representations of the gauge group, the spectrum of orientifold compactifications will also contain scalar fields in (anti)-symmetric or $({\mathbf{N}}_a, \mathbf{N}_b)$ representations of the gauge group. To study the influence of these fields on the existence of non-supersymmetric local minima, let us consider the same gauge group as before but additional charged scalars $\phi_{12^\star}$, $\phi_{3^\star 1}$ and $\phi_{23^\star}$ in $({\mathbf{N}}_a, \mathbf{N}_b)$ representations. Restricting the analysis to the case where the superpotential vanishes at the minimum, $\langle W \rangle$ = 0, some fields must be assumed to have a vanishing vev. Let us take for instance $\langle |\phi_{23}|^2\rangle = \langle |\phi_{23^\star} |^2\rangle = 0$. At this point of the moduli space, the potential writes, $|(W_{123}|^2=|W_{123^\star}|^2=h^2)$, 
\bea
V(\phi_i) &=&h^2|\phi_{12}|^2\left(|\phi_{31}|^2  + |\phi_{3^\star 1}|^2\right)  + \left(|\phi_{12}|^2+|\phi_{12^\star}|^2 -|\phi_{31}|^2-|\phi_{3^\star 1}|^2 + \xi_1\right)^2
\label{flavors:ScPot5}
\\
&+&
\left(-|\phi_{12}|^2+|\phi_{12^\star}|^2+|\phi_{23}|^2+|\phi_{23^\star}|^2 +\xi_2\right)^2 + \left(|\phi_{23^\star}|^2-|\phi_{23}|^2 +\xi_3\right)^2\, .
\nonumber 
\eea
Two cases arise, depending on the flavor spectrum. If there are flavors in all bifundamentals, it can be shown that the potential  (\ref{flavors:ScPot5}) does not possess any non-supersymmetric minima. Otherwise, minima can be found for particular choices of flavor spectra. For instance, if there are no fields in the $(\mathbf{N}_2,\mathbf{N}_3)$ representation, $I_{23^\star} = 0$, a non-supersymmetric vacuum exist  at points where $\langle |\phi_{12}|^2\rangle ,  \langle |\phi_{12^\star} |^2\rangle , \langle |\phi_{31} |^2\rangle \neq 0$ if the parameters $\{h^2, \xi_1, \xi_2, \xi_3\}$ are in a domain where
\be
D_2 = -D_1 > 0\quad , \quad  D_2 + D_3 <0 \quad , \quad D_2 - D_3 >0  \, .
\label{orient:condnonsusy}
\ee
 The remaining fields are stabilized at zero vev. The  $F$-term $F_{\phi_{23}}$  and all $D$-terms have a non-zero vev. Supersymmetry is then spontaneously broken, but the charged fields classically  have a positive mass. At the minimum  where
 \be
\langle D_2 \rangle = - \langle D_1 \rangle \quad , \quad h^2 \langle |\phi_{12}|^2\rangle = -2 \langle D_2 \rangle -2 \langle D_3 \rangle \quad , \quad h^2\langle |\phi_{31}|^2 \rangle = 4\langle D_2 \rangle
 \label{orient:minnonsusy}
\ee
the mass of the fields stabilized at zero vev are given by
\be
M^2_{\phi_{3^\star 1}} \sim \langle D_2 \rangle - \langle D_3 \rangle  \quad , \quad M^2_{\phi_{23}} \sim \langle D_2 \rangle - \langle D_3 \rangle \, .
\label{orient:mass}\ee
whereas the mass matrix of the Higgs fields reads
\be
M^2 (\phi_{12}, \phi_{12^\star}, \phi_{31}) = 
\left(
\begin{array}{ccc}
4\langle |\phi_{12}|^2\rangle & 0 & (h^2 -2)\langle \phi_{12}\phi_{31}^\dagger \rangle
\\
0 &4\langle |\phi_{12^\star}|^2\rangle & 2\langle \phi_{12^\star}\phi_{31}^\dagger \rangle
\\
(h^2 -2)\langle \phi_{12}^\dagger \phi_{31} \rangle& 2\langle \phi_{12^\star}^\dagger\phi_{31}\rangle &4\langle |\phi_{31}|^2\rangle  \, .
\end{array}
\right)
\label{orient:mass2}
\ee
Classically, the local minimum is therefore stable.

One may wonder if non-supersymmetic minima exist when the number of simple gauge groups is bigger than three. To answer that question, let us consider the same example as in section \ref{abelian}, namely a square of abelian gauge groups $G=U(1)\times U(1)\times U(1)\times U(1)$ with the chiral matter given in Table \ref{table:4abelian} with minimal flavor numbers $|I_{ab}|,|I_{ab^\star}|=0,1$. As done previously,  the analysis is restricted to the case where $\langle W\rangle =0$. For each triangle of gauge groups with  non-vanishing trilinear couplings,  at least one scalar must then have a vanishing vev. At this point of the moduli space, the scalar potential can for instance be written as
\bea
V(\phi_i) &=&h^2|\phi_{12}|^2\left(|\phi_{23}|^2 +|\phi_{41}|^2+|\phi_{23^\star}|^2 + |\phi_{4^\star 1}|^2\right)
\label{flavors:ScPot4}
\\
&+&
\left(|\phi_{12}|^2+|\phi_{12^\star}|^2 -|\phi_{41}|^2-|\phi_{4^\star 1}|^2  +\xi_1\right)^2+\left(-|\phi_{12}|^2+|\phi_{12^\star}|^2+|\phi_{23}|^2+|\phi_{23^\star}|^2 +\xi_2\right)^2 
\nonumber 
\\
&+&
\left(|\phi_{23^\star}|^2-|\phi_{23}|^2+|\phi_{34}|^2+|\phi_{34^\star}|^2 +\xi_3\right)^2+\left(|\phi_{34}|^2-|\phi_{34^\star}|^2+ |\phi_{4 1}|^2-|\phi_{4^\star 1}|^2 +\xi_4\right)^2 \, .
\nonumber 
\eea

Again, two cases arise, depending on the flavor spectrum.  If all flavor numbers  are chosen to be non-vanishing, it can easily  be shown that  the potential (\ref{flavors:ScPot4}) has no other extrema than the supersymmetric one. Otherwise, extrema are only found at points of the parameter  space $\{h^2,\xi_1,\dots, \xi_4\}$  for particular flavor configurations.  For instance, if there are no flavors in bifundamental representations
\be
I_{12^\star} = I_{24} = I_{3^\star 1} = 0 \, ,
\ee
there exists a minimum at points of the moduli space where
\be
\begin{array}{rcl}
\langle |\phi_{31}|^2\rangle = \langle |\phi_{34^\star}|^2\rangle = \langle |\phi_{23^\star}|^2\rangle =\langle |\phi_{41}|^2\rangle = \langle |\phi_{24^\star}|^2\rangle = 0 
\\ \langle |\phi_{12}|^2\rangle , \langle |\phi_{34}|^2\rangle , \langle |\phi_{23}|^2\rangle , \langle |\phi_{4^\star 1}|^2\rangle \neq 0 
\end{array}
\ee
if the parameters $\{ h^2, \xi_1,\dots , \xi_4\}$ are in the domain where
\be
\langle D_3 \rangle = \langle D_3 - D_2 \rangle = \langle D_2 \rangle > 0  \, .
\label{condnsusygen}
\ee
At these points, the classical mass matrix is a direct product of  the mass of scalars with zero vevs  
\be
\begin{array}{cc}
 M^2_{\phi_{34^\star}} \sim M^2_{\phi_{23^\star}} \sim \langle  D_3 \rangle 
 &\; , \;
 M^2_{\phi_{24^\star}} \sim  h^2 \langle |\phi_{4^\star 1}|^2 \rangle +  4 \langle D_3 \rangle
\\
  M^2_{\phi_{41}}  \sim  h^2 \langle |\phi_{12}|^2 \rangle +  4 \langle D_3 \rangle
&\; , \; 
  M^2_{\phi_{31}}  \sim  h^2 \langle |\phi_{23}|^2 \rangle +  4 \langle D_3 \rangle \; ,
  \end{array}
\ee
with the mass matrix of the Higgs fields $\{\phi_{12}, \phi_{34}, \phi_{23},\phi_{4^\star 1}\}$
\be
M_{{\rm Higgs}}^{2}  = 
\left(
\begin{array}{cccc}
4\langle |\phi_{12}|^2\rangle & 0 & (h^2 -2)\langle \phi_{12}\phi_{23}^\dagger \rangle &(h^2 -2)\langle \phi_{12}\phi_{4^\star 1}^\dagger \rangle
\\
0 &4\langle |\phi_{34}|^2\rangle & -2\langle \phi_{34}\phi_{23}^\dagger \rangle& -2\langle \phi_{12}\phi_{4^\star 1}^\dagger \rangle
\\
(h^2 -2)\langle \phi_{12}^\dagger \phi_{23} \rangle& -2\langle \phi_{34}^\dagger\phi_{23}\rangle &4\langle |\phi_{23}|^2\rangle & 0
\\
(h^2 -2)\langle \phi_{12}^\dagger \phi_{4^\star 1} \rangle &-2\langle \phi_{34}^\dagger \phi_{4^\star 1} \rangle&0 &4\langle |\phi_{4^\star 1}|^2\rangle
\end{array}
\right)\, .
\label{orient:mass2b}
\ee
It can be proven that all mass eigenvalues are positive when the parameters are in the non-supersymmetric domain (\ref{condnsusygen}).
This case may be easily generalized to any number of gauge factors with bifundamental matter. Locally stable vacua with supersymmetry  is therefore only possible for particular choices of flavors. In abelian cases, some flavors must be absent  in order  to obtain a supersymmetry broken vacuum.

Local non-supersymmetric minima are possible in particular points of the parameter space spanned by the trilinear coupling $h^2$ and FI-terms  $\xi_a$. Note that this domain does not overlap with the supersymmetric domain. Therefore, at a perturbative level, a theory with a given gauge and matter content can not have a supersymmetric and a non-supersymmetric vacuum at the same time. This will become obvious in string constructions where the parameter space is spanned by closed string moduli. Yet, the question of non-perturbative transition between these two vacua remains open.

\section{String Construction}\label{string}

The mechanism presented in section \ref{basic} can be implemented in  string constructions. Indeed magnetized branes in  Type IIB orientifold compactifications include its three essential building blocks: the gauge and matter spectrum, FI-parameters and Yukawa couplings. First,  stacks of magnetized $Dp$ branes with different magnetic fluxes on their worldvolume have the suitable spectrum. It leads both to unitary gauge groups and to massless chiral spinors either in bifundamental or (anti-)symmetric representations of the gauge group. In addition to these, there are scalars in the same representations whose mass generically depends both on the fluxes and on the volume moduli of the internal manifold. Then, FI-parameters in the $D$-term depend on the K\"ahler moduli. Trilinear couplings in the superpotential are functions of the complex structure moduli.  The FI-term can be computed by a supersymmetrization of the four-dimensional topological couplings \cite{Dine:1987xk,Atick:1987gy}.   Similarly, the Yukawa couplings  are obtained by the usual compactification of the brane action to four dimensions   \cite{Strominger:1985it}. This has been explicitly done for parallel fluxes \cite{Cremades:2004wa} in a toroidal compactification, while the most general case involving oblique fluxes has not been fully solved yet \cite{Bertolini:2005qh}.
These features arise with strong constraints coming from consistency conditions called RR-tadpole conditions. They ensure the finiteness of one-loop amplitudes and the cancellation of all anomalies. They then  restrict the rank of the gauge groups and the allowed matter content.

Let us be more precise and focus on toroidal   orientifold compactification of $K$ stacks of space-time filling $D9$ branes with a $U(1)_a$ gauge bundle on their worldvolume, $a=1,\dots, K$. Let us assume that the connections of these bundles have a constant field strength $F_a$ on the internal part of the world-volume. The gauge bundle is then characterized by some set of Chern numbers $m_a \in \mathbb{Q}$ in the internal directions. The corresponding boundary states can be written in terms of the rotation matrices $R_a$ \cite{DiVecchia:1999rh}
\be
R_a = \left(1-F_a\right)\left(1+F_a\right)^{-1} \, .
\label{ws:reflection}
\ee
When these matrices commute $[R_a,R_b]=0$, $\forall a,b=1,\dots, K$, the flux are usually called {\emph{parallel}}.   The T-dual configuration corresponds to intersecting $D6$ brane, where the fluxes are mapped under T-duality to the homology class the different $D6$ branes wrap \cite{Review}. In the more general case   where the rotation matrices do not commute, $[R_a, R_b]\neq 0$, the fluxes are called \emph{oblique}  \cite{AM, Bianchi:2005yz}. The T-dual configuration involves $Dp$-$Dq$ configuration of branes with magnetic fluxes on their worldvolume ("coisotropic" brane)\footnote{An example of $D6$-$D8$ system has been presented in \cite{Font:2006na}} \cite{Anastasopoulos:2006hn}.

Magnetized $D9$ branes in toroidal compactification of Type I string theory will be considered here.  The orientifold projection ${\cal O}=\Omega_p$ is defined by the worldsheet parity $\Omega_p$ and therefore  leave the ten-dimensional target-space invariant.
The associated orientifold planes are then space-time filling, giving rise exclusively to $O9$ planes. The closed string sector is ${\cal N}=4$ supersymmetric.  Even if  other choices of orientifold projections may lead to less conserved supercharges, we will however keep this easy compactification for the sake of simplicity. 

 \subsection{Open string spectrum}\label{sec:spec}
Only states which are invariant under the orientifold projection are kept in the physical spectrum. 
Since the world-sheet parity $\Omega_p$ maps the left moving into the  right moving sector, the boundary condition keeps this form, but with inverse reflection matrices $(R^{a})^{-1}$. Using the definition (\ref{ws:reflection}), the orientifold projection maps a flux ${\cal F}^{a}$ to its mirror flux $-{\cal F}^{a}$. 
Therefore, in order to have an invariant configuration of magnetized $D9$ branes in type I toroidal compactification, both fluxes ${\cal F}^{a}$ and $-{\cal F}^{a}$ must be present. The set of  stacks of branes must then be augmented. In addition to the K stacks of $N_a$ magnetized $D9$ branes, there must exist K stacks of $N_{a^\star}$ branes, with multiplicities $N_a = N_{a^\star}$ and fluxes ${\cal F}_{a^\star}=-{\cal F}_{a}$. 

The whole tower of twisted open strings spectrum is then affected. In addition to the sector of open strings located on a brane   (($aa$)-sector) and the open strings stretched between two branes (($ab$)-sector), two additional sectors appear. Together, we obtain
\begin{itemize}
\item {\bf $(aa)$-sector}: The massless states are organized in a ${\cal N}=4$, $d=4$ $U(N_a)$ vector multiplet. If the Chan Paton factors  are invariant under the projection ${\cal O}_{(1)}$, the gauge group is reduced to the orthonormal group $SO(2N_a)$ or symplectic group $USp(2N_a)$

\item {\bf $(ab)$-sector}: Its massless spectrum contains massless chiral $SO(1,3)$ fermions in the bifundamental representation $(\bf{N}_a, \bar{\bf{N}}_b)$ with some light scalars in the same representations. Their multiplicities are given by the intersection number $I_{ab}$ defined by
\be
I_{ab^\star} ={1\over (2\pi)^3} \int_{\Sigma_6}\left[c({\cal F}_a)\wedge c(-{\cal F}_b) \right]_{top}\, ,
\label{ws:intersection}
\ee

\item {\bf $(ab^\star)$-sector}: The chiral fermions are in the representation $(\bf{N}_a, \bf{N}_b)$ and its multiplicity is given by 
\be
I_{ab^\star} ={1\over (2\pi)^3} \int_{\Sigma_6}\left[c({\cal F}_a)\wedge c(-{\cal F}_b) \right]_{top}\, ,
\label{ws:interstar}
\ee

\item {\bf $(aa^\star)$-sector}: Contrarily to the last two cases, this sector is invariant under the orientifold projection. Only the Chan-Paton degrees of freedom are truncated.  Out of the $I_{aa^\star}$ chiral states, $I_{aa^\star}^+$ of them will be in the symmetric representation of $U(N_a)$ and $I_{aa^\star}^-$ are in the antisymmetric representation, where
\be
I_{ab^\star}^\pm ={1\over (2\pi)^3} \int_{\Sigma_6}\left(1\pm \Omega_p\right)\left[c({\cal F}_a)\wedge c(-{\cal F}_b) \right]_{top}
\ee
\end{itemize}

\subsection{FI-terms }\label{sec:FI}

Magnetic fluxes on the worldvolume of the $Dp$ branes generate a FI-term in the scalar potential that can be computed by the supersymmetrization of the  four dimensional topological couplings  of the Wess-Zumino action \cite{Dine:1987xk,Atick:1987gy}. It  depends on both the flux quanta $F_a$ and the K\"ahler moduli $J$. In Type I compactification in particular,  magnetized $D9$-branes generate for each $U(1)_a$ gauge components with gauge couplings $g_a$ a FI-term $\xi_a$
\be
\label{xiST}
{\xi_a \over g^a_2} = 
M_s^2{1\over (4\pi^2\alpha')^3} \int\left( {\cal F}\wedge {\cal F}\wedge {\cal F} - J \wedge J \wedge {\cal F}  \right) 
 \, 
\ee
where ${\cal F}_a = 2\pi\alpha' F_a$ and $\varphi$ is the dilaton.  In the four-dimensional quadratic effective action, FI-term $\xi_a$ enters in the D-term with charged fields $\phi_i$ associated to the lowest lying open string states charged under the $U(1)_a$ gauge group. The auxiliary field $D$  then reads 
\be
D_a = \sum_i q^a_i |\phi_i|^2 K_{ii}(\phi,\bar{\phi})  + \xi_a \, ,
\label{Ddef}
\ee
where $K_{ii}$ is the quadratical term of  the K\"ahler potential for charged fields
\be
{\cal K }(\phi, \bar{\phi}) = \sum_{ij} K_{i\bar{j}}(J, \tau,\varphi)\phi_i \bar{\phi}_{\bar{j}}\, .
\label{string:kahlerpot}
\ee
It depends on the dilaton field $\varphi$, complex structure $\tau$ and twist angles $\theta_r^i$ as \cite{Review}
\be
K_{i\bar{j}}(J, \tau,\varphi) = \delta_{i\bar{j}}\; e^{-\varphi}V_6\prod_{r=1}^3
J_r^{-\theta_r^{i}/2}
\left({\rm   Im }\;\tau\right)^{\theta^{i}_r-\beta}\sqrt{{\Gamma(\theta_r^{i})\over\Gamma(1-\theta_r^{i}) }}\, ,
\ee
 where $\Gamma(\theta)$ denote the usual Gamma-Functions and $\beta$ a rational number. The quadratic approximation neglecting higher powers of the charged fields in the K\"ahler potential (\ref{string:kahlerpot})  is however only valid for small field values $v_i^2 := \langle|\phi_i|^2\rangle\ll M_s^2$.  

In absence of any vev for the open string fields $v^2_i \equiv 0$,  supersymmetry conditions $\xi_a=0$ are in agreement with the calibration condition given by \cite{Marino:1999af}, 
\be
{\rm Im}\; e^{i\theta_a}\left( iJ+{\cal F}_a\right)^3 = 0 \quad \quad , \quad \quad F^a_{2,0}=0 \, , 
\label{mmms}
\ee
for  $\theta_a$-parameter $\theta_a = -{\pi \over 2}$, $\forall a=1,\dots, K$. The second set of conditions $F_{2,0}^a=0$ arises from the F-flatness conditions of the geometrical moduli of the $Dp$-branes. The charged fields $\phi_i$ are then massless and form, together with a massless chiral fermion, a ${\cal N}=1$ chiral multiplet. Moreover, the gauge couplings $g_a^2$ have a polynomial form as the FI-term. It reads
\be
{1\over g_a^2} = e^{-\varphi}{1\over (4\pi^2\alpha')^3} \int\left( J\wedge J\wedge J - J \wedge {\cal F} \wedge {\cal F}  \right) \, .
\label{ga2}
\ee
More generally, for non-zero vev $v_i^2\neq 0$,  the polynomial form of the FI-term and gauge couplings remains valid in the quadratic approximation $v^2_i \ll M_s^2$. The supersymmetry conditions $D_a=0$ relate then the vev $v_i^2$ of the canonically normalized fields  $\tilde{\phi}_i$ with the fluxes quanta and K\"ahler moduli
\be
\langle D_a \rangle =0 \quad : \quad \sum_i q^a_i \langle |\tilde{\phi}_i|^2 \rangle+ \langle \xi_a({\cal F}_a , J)\rangle = 0
\label{string:Dflat}
\ee
The charged fields obtain a positive mass from the fluxes. Since the gauge symmetry is spontaneously broken by the Higgs phase of the charged fields $\phi_i$, the $U(1)$ vector bosons also become massive. Then these form the bosonic part of a massive vector multiplet. More precisely, not only the charged scalar get stabilized but also a combination of the K\"ahler and open string moduli. We will come back to this issue in section \ref{ScPot}.

\subsection{Yukawa couplings }\label{sec:YUK}

Let us now analyze the possible perturbative terms in the charged fields' superpotential. The holomorphicity and gauge invariance strongly restrict terms that may arise from magnetized branes. From the spectrum given in section \ref{sec:spec}, one deduces that  linear and quadratic terms are absent. Assuming the presence of at least three magnetized $Dp$-branes in Type I string theory, the first possible contributions are trilinear terms
 \be
 W^{(3)} = \sum_{ijk} W_{ijk} {\rm Tr} \; \phi^i_{ab}\phi^j_{bc}\phi^k_{ca} + \sum_{ijk}W_{ijk}{\rm Tr} \; \phi^i_{ab}\phi^j_{ba^\star}\phi^{k\; \pm}_{a^\star a} \,
 \ee
 where in the second term, the bifundamental fields $\phi^i_{ab}$ and $\phi^j_{ba^\star}$ must be correctly (anti-) symmetrized, depending on the representations of the field $\phi^{k\; \pm}_{aa^\star}$.  For systems of magnetized $D9$-branes with {\it parallel} fluxes, trilinear couplings depend on the Theta-functions as  \cite{Cremades:2004wa}
 \be
 W^{ijk}(\tau) =  \prod_{r=1}^3\theta \left[\begin{array}{c} \delta_{ijk}^r
 \\
 0\end{array}\right] (0,\tau^{(r)}| I^{r}_{ab}I^{r}_{bc}I^{r}_{ca}|) \, 
 \label{string:supot}
 \ee 
 where the indices $i,j,k$ indicate the flavor dependance and 
 \be
 \delta^{r}_{ijk} = {i^{(r)}\over I_{ab}^{(r)}}+{j^{(r)}\over I_{ca}^{(r)}}+{k^{(r)}\over I_{bc}^{(r)}} \quad \quad ; \quad \quad i,j,k^{(r)}=0,\dots,|I_{ab}|^{(r)}-1
 \ee
We then see that for each non-trivial triangle $(a,b,c)$ where $I_{ab}^{(n)},I_{bc}^{(n)},I_{ca}^{(n)} \neq 0 $, there will be a  trilinear coupling in the superpotential that depends on the complex structure of the internal torus. The same is expected to be true for  {\it oblique} fluxes, even if its generalization  is not yet known.

 Let us apply the superpotential (\ref{string:supot}) to the structure of the ${\cal N}=1$ vacua of magnetized $D9$ branes. We analyze magnetized branes in a Higgs phase, namely when some of the matter scalars, either in (anti-)symmetric or the bifundamental representation, acquire a non-trivial background value. Aside from the $D$-flatness conditions discussed in section \ref{sec:FI}, there is a second set of conditions for the existence of supersymmetric vacua that come from the F-flatness conditions
\be
\langle F_i \rangle := \langle {\cal D}_{\phi_i}W \rangle=0  \, 
 \ee
where ${\cal D}_{\Phi_i}$ denotes the covariant derivative in respect to the chiral fields $\Phi_i$, $ {\cal D}_{{\phi_i}} = \partial_{\phi_i} + \partial_{\phi_i} {\cal K}$. In our analysis, we will focus on the stronger Minkowskian constraints 
 \be
 \partial_{\phi_i} W = W =0\, .
 \label{string:Fflat}
 \ee 
This implies that for each "triangle" at least two fields must have a zero vev in order to form a supersymmetric vacuum. In addition to Yukawa couplings, trilinear terms in the superpotential give rise quartic scalar couplings. Therefore, even if the $F$-flatness conditions are satisfied, the charged fields cannot  be stabilized only by this $F$-term. A mass term can only be lifted if additional terms in the scalar potential are added. Since $\mu$-terms are absent of the perturbative superpotential, combined effect of trilinear terms in the superpotential and FI-terms in the $D$-terms will be analyzed here.

  \subsection{Tadpole conditions}\label{sec:tad}

In compactifications on compact manifolds, the conditions presented in sections \ref{sec:FI} and \ref{sec:YUK} are not sufficient for supersymmetry. In a global construction, additional  conditions must be imposed to obtain consistent supersymmetric models. Indeed,  magnetized branes induce tadpole couplings depending on the flux quanta in the internal space and the rank of the gauge group of the different stacks. For compact internal manifolds, the sum of the overall RR-tadpole charges must vanish. The possible flux configuration and gauge group are then restricted. It ensures in particular that the spectrum is  anomaly-free. In type I string theory, these conditions read \cite{Bianchi:2005sa}
 \be
 \sum_a N_a{\rm det}W_a  = 16 \quad , \quad \sum_a N_a{\rm det}W_a \int_{\Pi_4^A} F_a\wedge F_a  =  q_{\;\; O5}^{A}
\label{TO9}
\ee
where ${\rm det}W_a$ denotes the winding matrix of the brane over the torus.  In orientifold compactification considered here,   5-brane charges $q_{O5}^A$ all vanish since the orientifold projection leaves the ten-dimensional spacetime invariant.

\section{Scalar potential from $F$ and $D$-terms}\label{ScPot}

In  section \ref{string}, it has been shown how magnetized branes naturally contribute to the superpotential  and $D$-term of their four-dimensional effective action. We want to apply these results to a new mechanism to generate a potential and stabilize the  twisted open string moduli. As it has been shown in section \ref{basic}, the combined effects of $F$-term and $D$-term to the scalar potential give rise to local minima for the charged fields that may be either supersymmetric or non-supersymmetric, depending on the values of the FI-parameters  $\xi_a(J,{\cal F}_a)$ and trilinear couplings $W_{ijk}(\tau)$. Moreover, in presence of multiple stacks of branes with bifundamental or (anti)-symmetric scalars at their intersection, the consistency of $D$-flatness conditions of the different branes also  restrict the possible values of the FI-terms and gives a mass term for the K\"ahler moduli. 

 In addition to these recombination fields, there also exists adjoint scalars from the lowest excitation of the untwisted sector that enters in the ${\cal N}=4$, $d=4$ $U(N_a)$ vector multiplet in toroidal compactification. Their stabilization goes beyond the scope of this paper. However, they can be explicitly projected out from the massless spectrum in compactifications where massless string excitations in the untwisted open string sector preserves ${\cal N}=1$ supersymmetry, e.g. in Calabi-Yau compactification. Note that since $D9$ branes cover the entire ten-dimensional space, there are no geometric moduli that parametrize their position in the internal space. For lower dimensional branes however, their stabilization  has been treated in \cite{Angelantonj:2003zx, Lust:2005bd,Gomis:2005wc}.

Let us be more specific and restrict our analysis to the case where the value of the superpotential at the vacuum vanishes $\langle W=0 \rangle $. The FI-term (\ref{xiST}), the superpotential (\ref{string:supot}) and the K\"ahler potential (\ref{string:kahlerpot})   lead to the scalar potential for the unnormalized charged fields of the form 
\be
\begin{array}{rcl}
V(\phi_i, J,\tau,\varphi) &=&\sum_i (K^{-1})^{ii}(J,\varphi)F_i(\tau,\phi_i) F_i^\star(\tau,\phi_i) + \sum_a g_a^{-2}\,\left(\sum_i q_i^aK_{i\bar{i}}|\phi_i|^2  +\xi_a(J,{\cal F}_a)\right)^2
\\
&= &(K^{-1})^{i\bar{i}}\left(W_{i}^{\; jk}W_{i}^{\star \; mn} \phi_j\phi_k\phi^\star_m\phi^\star_n \right)+ \sum_a\,g_a^{-2}\,\left(\sum_i q_i^aK_{i\bar{i}}|\phi_i|^2  +\xi_a(J,{\cal F}_a)\right)^2
\end{array}
\label{string:ScPot}
\ee

\subsection{Supersymmetric minima}\label{ScPotSUSY}

Since both $F$ and $D$-term contributions to the scalar potential (\ref{string:ScPot}) are positive, the global minimum sits at the supersymetric points $\langle F_i\rangle=\langle D_a\rangle=0$. In type II compactification,  for all "triangles "  of branes $(a,b,c)$ where $W_{ijk}\neq 0$, the $F_i$-flatness condition $\langle F_i\rangle=0$ restricts the moduli space  to point where all charged fields but one must have a vanishing vev, while they do not constraint the remaining one. $D$-flatness conditions $\langle D_a\rangle=0$ related the  vev  of the unconstrained fields to the value of the FI-parameters, and consequently to the K\"ahler moduli. However, as in the toy models presented in section \ref{basic}, consistency of the supersymmetric conditions of different abelian factors restricts the possible FI-parameters. Then, if the number of stacks is greater than the number of charged fields unconstrained by the $F$-flatness, $D$-terms also  stabilize the K\"ahler moduli, as in the mechanism presented in  \cite{AKM2}.

For the unnormalized charged fields whose vev is fixed at the origin by $F$-term, the mass
\be
\langle\phi_J\rangle = 0 \; : \;
M^2_{\phi_J} = (K^{-1})^{i\bar{i}} \left(\sum_{kn}W_{i}^{\; Jk}W_{i}^{\star \; Jn} \langle \phi_k \rangle\langle\phi_n^\star\rangle\right) 
\label{massphi0}
\ee
 depend on the vev  $\langle \phi_k\rangle$.  For exclusively trilinear terms  in the superpotential,   the $F$-term  is therefore not sufficient to stabilize the moduli. Combined effects of $F$ and $D$-terms must be considered. $D$-term contributions determines the vevs of the unconstrained fields in terms of the FI-terms and consequently in terms of the K\"ahler moduli.   
 
 Taking into account that at most one field in each "triangle" acquires a non-vanishing vev in a supersymmetric vacuum, the mass of the correctly normalized fields $\tilde{\phi}_J$ 
\be
M^2_{\tilde{\phi}_J} = (K^{-1})^{J\bar{J}} \left(\sum_{i\, ,n}(K^{-1})^{i\bar{i}}(K^{-1})^{n\bar{n}}W_{i}^{\; Jn}W_{i}^{\star \; Jn} \langle |\tilde{\phi}_n|^2 \rangle \right)
\label{massphi0Bis}
\ee
 is a function of the vev of the metric moduli via the trilinear couplings and K\"ahler potential.

On the other hand, the mass of Higgs  fields does not depend on the trilinear couplings. Indeed, since on each "triangle" only one field may acquire a non-vanishing vev without breaking the $F$-flatness,  the mass of the normalized field whose vev is $\langle |\tilde{\phi}_k|^2 \rangle = -\xi_a / q_k^a$ reads 
\be
 M^2_{\tilde{\phi}_k} \sim \langle|\tilde{\phi}_k|^2\rangle
 \sum_{b}(q_k^b)^2
  \quad {\rm and} \quad
 \langle|\tilde{\phi}_k|^2\rangle = -{\xi_a \over q_k^a} \quad , \forall a \; {\rm where}\quad  q_k^a\neq 0  \, .
 \label{massphiNot0}
 \ee
 Two separate cases arise depending on their charges $q_k^a$.  For bifundamental $\phi_{ab}$ and (anti)-symmetric $\phi_{aa^\star}$ fields,  we then obtain
 \be
 \begin{array}{rcl}
 \phi_{ab} &:& M^2_{\tilde{\phi}_{ab}} \sim 2\langle|\tilde{\phi}_{ab}|^2\rangle
  \quad {\rm where} \quad
 \langle|\tilde{\phi}_{ab}|^2\rangle = -\xi_a =\xi_b
 \\
  \phi_{aa^\star} &:& M^2_{\tilde{\phi}_{aa^\star}} \sim 4\langle|\tilde{\phi}_{aa^\star}|^2\rangle
  \quad {\rm where} \quad
 \langle|\tilde{\phi}_{ab}|^2\rangle = -\xi_a / 2 \, .
\label{bifund+antisym}
 \end{array}
 \ee
  In orientifold compactifications  where the intersection numbers $I_{ab}$ and $I_{ab^\star}$ do not vanish, there are bifundamental fields both in $(\mathbf{N}_a,\bar{\mathbf{N}}_b)$ and $(\mathbf{N}_a,\mathbf{N}_b)$ representations. $F$-terms then leave two fields  unconstrained, as explained in section \ref{flavors}. Their mass and vev are given in terms of the FI-terms $\xi_a$ and $\xi_b$ as
 \be
\langle|\tilde{\phi}_{ab}|^2\rangle = \xi_b - \xi_a \; , \; \langle|\tilde{\phi}_{ab^\star}|^2\rangle = -\xi_b - \xi_a  \quad : \quad
 M^2_{\tilde{\phi}_K} \sim 2\langle|\tilde{\phi}_K|^2\rangle
 \label{vevphiNot0}
 \ee
 
 In the quadratic approximation, the vevs must be small in comparison with the string scale. Consequently, the scalar mass (\ref{massphiNot0}) of the Higgs fields must be much smaller than the string Mass, $M_{\tilde{\phi}_K}\ll M_s$. Moreover, since the trilinear couplings (\ref{string:supot}) depend exponentially on the complex structure, there may exist a second hierarchy of masses. Indeed, points of the closed string moduli space exist where the trilinear couplings and K\"ahler potential are very small.  In these points, the masses (\ref{massphi0Bis}) will be hierarchically smaller than the Higgs masses (\ref{massphiNot0}).

Most of the recombination fields are therefore stabilized by the combined effects of $F$- and $D$-terms. As explained in section  \ref{basic}, the mechanism has some limitations. On one hand,  a full stabilization is only achievable if the number of flavors  fixed by $D$-term is minimal, i.e $I_{ab}=1$. Indeed, each $D$-flatness condition restrict a single field for each $U(1)_a$ gauge factor. Supplementary chiral fields in this family  will therefore be flat directions of the scalar potential. On the other hand, this method is restricted to fields charged under the abelian subgroup of the gauge symmetry.  Indeed, in order to stabilize some scalars at non-zero vevs, it is crucial to have constant terms in the $D$-terms. Positive masses from $D$-terms are therefore only possible for non-vanishing FI-terms. This can only be implemented for abelian factors of the gauge symmetry. 

K\"ahler moduli can also be stabilized by $D$-terms. As it has been shown in section \ref{basic},  $D$-flatness conditions arising from different $U(1)$ factors can only be consistent if the FI-terms $\xi_a$ are restricted in some particular domains. Since they depend on flux quanta and K\"ahler moduli, these restrictions can be interpreted as a K\"ahler moduli stabilization. The stabilization mechanisms works as follows. The $F$ and $D$-flatness conditions stabilize most of the open string moduli. If the number of fields charged under the abelian groups is smaller than the number of $U(1)$ factors, $D$-flatness conditions will constrain the K\"ahler moduli via the FI-terms. For instance, in the case of three magnetized brane with antisymmetric matter, the FI-term of one of the abelian factor must vanish. A least a K\"ahler modulus is then stabilized. 

\subsection{Non-supersymmetric Vacua}\label{ScPotNSUSY}

One may ask if the scalar potential (\ref{string:ScPot}) possesses other minima for the charged fields than the global minimum  studied in section \ref{ScPotSUSY} . These local minima arise at points of the field space where some $F$ and $D$-terms would not vanish. Supersymmetric is then broken. It is possible when the system of equations formed by $F$- and $D$-flatness conditions is overconstrained. This would happen when the number of unconstrained matter fields and K\"ahler moduli is smaller than the number of $U(1)_a$ factors. 

As shown in section \ref{basic}, there is a domain of the parameter space   where  non-supersymmetric minima for all charged fields exist. For instance, in the case of a single "triangle" of three abelian gauge group, the domain is given by eq (\ref{NSusyCond}) and relates the Yukawa couplings to the FI-terms. In string compactifications, these parameters depend on the closed string moduli. For certain regions of the closed string moduli space, there may exist classically stable non-supersymmetric vacua. 

This situation can be explicitly  be implemented by two sets of stacks of magnetized branes that can be called   {\it hidden} and   {\it visible} sector. In the first sector, supersymmetry is preserved. The number of branes is large enough to stablize all its matter and metric moduli by the mechanism presented in section \ref{ScPotSUSY}. The tadpole conditions are however not satisfied. A second sector must be added such that the total  RR charges vanish. Its gauge symmetry and matter content  is chosen in such a way that the scalar potential possesses only a local non-supersymmetric minima. It may be achieved only if the closed string moduli are stabilized by the first sector in a suitable domain. In the {\it visible} sector, supersymmetry is then spontaneously broken by a combined $F$- and $D$-terms and all charged scalars acquire a positive mass.

This forms a consistent construction when RR-tadpole conditions are satisfied and therefore all anomalies cancelled. However, since supersymmetry is broken,  a non-vanishing  NSNS-tapdole remain. Contrarily to the superpotential, $\langle W \rangle$=0,  some $F$ and $D$-terms have a non-zero vev.  The value of the scalar potential (\ref{string:ScPot}) at the minimum is then strictly positive.  Once the massive K\"ahler and complex structure deformations are integrated out, the scalar  potential becomes an effective potential for the dilaton. Since the Yukawa coupling and $D$-terms arise from disk amplitudes, it has a runaway behavior,  $V_{{\rm min}} (\varphi) = V_0 e^{-\varphi} >0 $. The equation of motion for the dilaton field can therefore not  be satisfied at finite values. In cases where the backreaction of the dilaton tadpole has been analyzed, it was found that the constructions in type II string theory stabilize the dilaton at strong coupling \cite{Blumenhagen:2000dc,Dudas:2000ff,Angelantonj:2007ts}. One may ask under what conditions could the dilaton be fixed at weak coupling. Solutions to this problem will be adressed in a later work  \cite{ADM}.

\section{Supersymmetric model}\label{sec:susy}
 
 We present in this section a model of magnetized $D9$ branes in  toroidal compactification of Type I string theory where a global ${\cal N}=1$ supersymmetry is conserved. We do not aim at presenting here fully realistic model whose low energy behavior reproduces the standard model. We will show the existence of consistent compactifications with some closed and open string moduli stabilized in a supersymmetric configuration. To achieve that aim,  the magnetic fluxes and the rank of the Chan-Paton matrices are chosen  to satisfy  the RR-tadpole conditions. In absence of any orbifold projections, the 5-brane charges induced by the different magnetized branes must add to zero, while their 9-brane charges must add to 16. 
 
This determines the rank of the gauge groups and the family replication of the matter content as explained in section (\ref{sec:spec}). Here, six stacks with parallel fluxes $F_i^a=F_{x_iy_i}^a$ are introduced as given in  Table \ref{table:susy1}. Their gauge symmetry is a product of six abelian factor plus $SU(2)\times SU(4)$ gauge group with chiral matter fields $\phi^k_{ab}$ and $\phi^k_{ab^\star}$ in bifundamental representations of each "intersections" $(ab)$ and $(ab^\star)$.  Charged fields  in antisymmetric representations appear only in  stack with $SU(4)$ gauge group, $\phi^{k}_{\;6\, 6^\star}$, where $k$ denotes here the flavor index of each chiral family. Six additional stacks of  branes have oblique fluxes identical to  the one of \cite{AKM2}. They give rise to six abelian gauge groups without chiral matter. 

It can  easily  be seen that the twelve stacks given in Table \ref{table:A} and \ref{table:susy1} satisfy all RR-tadpole conditions. One can then analyze the associated superpotential and D-term and look for supersymmetric minima. The holomorphic variables are chosen such that the trilinear couplings in the superpotential reads 
\bea
W &=&  \sum_{a=3}^{6}\sum_{ij} W^{ij}_{12a}\; \phi_{12}\, \phi^i_{2a}\, \phi^j_{a1} + \sum_{a=3}^{6} \sum_{ij} W^{ij}_{12a^\star}\; \phi_{12}\, \phi^i_{2a^\star}\, \phi^j_{a^\star1} 
\\
&+& \sum_{b=5}^{6} \sum_{ij}W^{ij}_{34b}\; \phi_{34}\, \phi_{4b}\, \phi_{b3} + \sum_{b=5}^{6}  \sum_{ij} W^{ij}_{34b^\star}\; \phi_{34}\, \phi_{4b^\star}^i\, \phi^j_{b^\star 3} + \sum_{ijk}W^{ijk}_{566^\star}\;  \phi^i_{56}\,\phi^j_{5^\star 6} \,\phi^k_{66^\star } \; , 
\eea
where the sum over $i,j,k$ run over the flavor indices. Since the intersection numbers $I_{12}$ and $I_{34}$ are equal to one, the chiral fields $\phi_{12}$ and $\phi_{34}$ do not cary any flavor indices. The couplings $W^{ij}_{1,2,3,4}$ are given in eq (\ref{string:supot}).  In addition to the complex structure, these also depend to the relative Chern numbers  on each triangle.

 \begin{table}
\begin{center}
\begin{tabular}{|c||c|c|c|c|c|c|}
\hline
 & $\sharp 1$  &$\sharp 2$  & $\sharp 3$& $\sharp 4$&$\sharp 5$  & $\sharp 6$
 \cr
 \hline
   \hline
  $N_a$& $1$  &$1$  & $1$& $1$&$2$  & $4$
 \cr
 \hline
 $(F_{1},F_{2},F_{3})$ & $(8,5,1)$  & $(7,2,0)$ & $(4,5,3)$&$(3,4,2)$ & $(0,12,-18)$&$(-1,17,6)$
\cr
\hline
\end{tabular}
\end{center}
\caption{Set of consistent  branes with parallel fluxes. They are characterized by the rank $N_a$ of their unitary gauge group  and by their flux quanta $F_i^a$ in the directions $(x_iy_i)$ of the torus. In third line are given the vev of the charged field at the minimum in string units. Note that the winding matrix is taken to be the unit matrix}
\label{table:susy1}
\vskip .4cm
\end{table}

The $F$-flatness conditions $F_{\phi_{ab}}=0$ and $F_{\phi_{ab^\star}}=0$ (  at zero superpotentiel $W=0$ )  for the chiral fields $\phi_{ab}$ and $\phi_{ab}$ restrict  all scalars but three to have zero vev. They however acquire  a mass from the F-term potential only if the unconstrained scalars $\phi_{12}$,  $\phi_{12^\star}$,  $\phi_{34}$, $\phi_{34^\star}$ and $\phi_{66^\star}^k$ possess a non-vanishing vev. Indeed, their mass reads
 \be
 \begin{array}{c}
 M^2_{\phi_{1a}}\sim M^2_{\phi_{a2}}\sim M^2_{\phi_{2a^\star}}\sim M^2_{\phi_{a^\star 1}}\sim \langle |\phi_{12}|^2 \rangle  
 \\
 M^2_{\phi_{4a}}\sim M^2_{\phi_{a3}}\sim M^2_{\phi_{4a^\star}}\sim M^2_{\phi_{a^\star 3}}\sim \langle |\phi_{34}|^2 \rangle  
  \\
 M^2_{\phi_{56}}\sim M^2_{\phi_{5^\star 6}}\sim \langle |\phi^k_{66^\star}|^2 \rangle   \, .
 \end{array}
 \ee
 We are therefore left with five unconstrained charged fields. These can be in turn stabilized by the $D$-flatness condition arising from the six abelian factor of the six branes with parallel fluxes. They read
 \be
 \left\{
 \begin{array}{l}
 D_1 = (|\phi_{12}|^2 + |\phi_{12^\star}|^2 + \xi_1) \, , \, D_2 = (-|\phi_{12}|^2 +|\phi_{12^\star}|^2 + \xi_2)  \\\
 D_3 = (|\phi_{34}|^2 +|\phi_{34^\star}|^2+ \xi_3)  \, , \, D_4 = (-|\phi_{34}|^2 +|\phi_{34^\star}|^2 + \xi_4) 
 \\
 D_5= \xi_5 \, , \, D_6 = 2\sum_k|\phi^k_{66^\star}|^2 + \xi_6
\end{array} 
\right.
\label{ex:susyD}
\ee
 The supersymmetry conditions $D_a=0$, $\forall a=1,\dots,6$ can be simultaneously satisfied if and only if the open string moduli and FI-term $\xi_a$ are constrained to be in the domain where
 \be
  \begin{array}{l}
 v_{12}^2 \equiv|\phi_{12}|^2 =\xi_2- \xi_1 \; ,\;   v_{34}^2 \equiv|\phi_{34}|^2 =\xi_4- \xi_3
 \\
 v_{12^\star}^2 \equiv|\phi_{12^\star}|^2 =-\xi_2- \xi_1 \; ,\;   v_{34^\star}^2 \equiv|\phi_{34^\star}|^2 =\xi_4- \xi_3
 \\
 \xi_5=0 \quad {\rm and} \quad v_6^2\equiv |\phi^k_{66^\star}|^2 =- \xi_6\, .
\end{array}
 \ee
 The FI-terms $\xi_a(F^a,J)$ depend on the flux $F^a$ on the brane and the Kahler moduli $J$. Since the off-diagonal K\"ahler moduli are stabilized by the $D$-term of the stacks with oblique fluxes, there are three unconstrained K\"ahler moduli describing the volume of the two-tori \cite{AM,AKM2}. The supersymmetry conditions \ref{ex:susyD} form a system of six equations for eight variables $v_{12}^2, v_{34}^2, v_6^2$ and $J_{1}, J_2, J_3$. It turns out that for the choice of fluxes given in Table \ref{table:susy1}, there exists two flat directions corresponding to the volume of two of the three two-torii. Only a single K\"ahler modulus is stabilized by $D$-term.  In the open string moduli space, there are therefore flat direction only in the direction of the antisymmetric scalars $\phi_{66^\star}^l$ that are not fixed by the $D$-flatness condition of the $U(1)_6$ gauge factor. 
 
 This model is particular in many aspects. First, the number of families in sectors $(12)$ and $(34)$ has been chosen to be minimal. If this would not be the case, only a single field would be stabilized, while the rest would remain massless.  Moreover, the abelian gauge groups has played a central role. A non-vanishing FI-parameters $\xi_a$ could then be introduced. This allowed then the charged fields to acquire a positive mass. For more general gauge groups, FI-term can only be introduced in Cartan subalgebra. Therefore, only fields charged under the Cartan subgroup of some gauge group may be stabilized at non-zero vev. Another consequence of the abelian gauge group is the absence of fields in antisymmetric representations. The number of charged fields unconstrained by $F$-flatness conditions has then been reduced from six to three, allowing in turn to stabilize both open string and K\"ahler moduli.

\section{Conclusions}

In this article, magnetized branes in Type I compactifications have been considered.  The compatibility between local conditions of supersymmetry and the global issues of tadpole conditions is analyzed.  Both conditions are incompatible in toroidal compactification. Indeed, the supersymmetry conditions in  is a calibration condition on the worldvolume geometry of the different magnetized branes, which can not be satisfied simultaneously with the cancellation of all 5-brane charges induced by the branes. 

But interpreted as a $D$-flatness condition, the calibration condition can be generalized to cases where charged fields are present. When some of them acquire a non-vanishing vev, the supersymmetry condition is modified, whereas the tadpole conditions remain the same. It is shown that in the quadratic field theoretical approximation, both local and global conditions can be satisfied simultaneously. 

Arguments for the vacuum  stability have been given. In usual compactifications, the charged scalars describe flat directions of the scalar potential. Their vevs are therefore not determined. For magnetized branes however, non-trivial perturbative potential can be generated from $F$ and $D$-term contributions. Indeed, trilinear couplings of the superpotential imply the presence of quartic couplings for the charged scalars. Together with the $D$-term contributions in presence of non-vanishing FI-term, they form a potential that possesses local minima. Under minimal assumptions on the spectrum, global supersymmetry  exists in principle with most of the charged scalars stabilized.  Some of them are stabilized at points of the moduli space with  non-vanishing vev. Gauge symmetry is therefore spontaneously broken with a light Higgs. Furthermore, a hierarchy of scalar masses can be naturally implemented by the exponential difference between the $F$-term couplings and FI-terms. 

It has then been shown what conditions must be set on the spectrum so that  metric moduli are also stabilized by magnetized branes. Once the charged fields are stabilized, K\"ahler moduli are in turn stabilized by $D$-terms. Complex structure moduli  also acquire a positive mass by $F$-terms. In principle, metric moduli and charged fields can be perturbatively stabilized in a Minkowski space. An explicit toroidal example with chiral spectrum has been provided where a supersymmetric vacuum exists. There, most of  the charged fields and some K\"ahler moduli are fixed by the combined effects of $F$ and $D$-terms. 

Beyond the global minimum, the existence of local non-supersymmetric vacua has been analyzed. It was shown that there are points of the metric moduli space for which  non-supersymmetric vacua exist and where the charged scalars have positive classical mass. These constructions are free of open string tachyons and therefore locally stable. However, even for constructions where all RR tadpoles are cancelled, there remains a disk tadpole for the dilaton field. One must therefore go beyond orbifold compactification to construct a non-supersymmetric locally stable theory. Effects arising from other orders in the string perturbation theory must be added  to be able to solve the equation of motion of the dilaton. This problem will be addressed in a later work \cite{ADM}. 

The theories with supersymmetric vacua are different from the ones with non-supersymmetric vacua. Their Chan Paton factors and flux quanta differ a priori. Therefore,  they have different gauge groups and different matter content. It would be of great interest to understand the transition via nucleation of branes from one vacuum to the other, as in \cite{Angelantonj:2007ts}.

One has also gone a step further in the understanding of the Higgs sector of Type I string theory. One has proposed a mechanism to generate  a scalar potential for the Higgs fields  from both $F$ and $D$-terms leading to non-trivial minima.   It would be interesting to implement this method in more realistic models in order to have a more predictive scenario from string theory for the sector of  spontaneous breaking of gauge symmetry. 

In a more conceptual viewpoint, a deeper analysis of the charged scalar's K\"ahler potential is essential. The stabilization mechanism presented here relies on its quadratic approximation. A true identification of the canonical coordinates of the open string and K\"ahler moduli space and their stabilization imply a knowledge of K\"ahler potential beyond this approximation. 

 \section*{Acknowledgements}
 
We would like to thank C. Angelantonj, I. Antoniadis, P. Camara, E. Dudas, V. Niarchos, F. Nitti, F. Villadoro for useful discussions. We also would like to thank J. Baranes for the careful reading of this manuscript.  This work was also partially supported by
INTAS grant, 03-51-6346, RTN contracts MRTN-CT-2004-005104 and
MRTN-CT-2004-503369, CNRS PICS \#~2530,  3059 and 3747,
and by a European Union Excellence Grant,
MEXT-CT-2003-509661.

\appendix

\section{Table A}\label{table:A}

\begin{center}
$$
\begin{array}{|c||c|c|c|}
\hline
\mathrm{Stack} \sharp & \mathrm{Multiplicity} &\mathrm{Fluxes}  & 
\mathrm{5-brane\ localization}\\
\hline
\hline
&&&\\
\sharp 1  & N_1=1&(F^1_{x_1 y_2},F^1_{x_2 y_1})=(1,1)
                                    &  [x_3,y_3]          \\
&&&\\
\hline
 &&&\\
\sharp 2  & N_2=1& (F^2_{x_1 y_3},F^2_{x_3 y_1})=(1,1)& 
  [x_2,y_2]\\
       &&&\\
\hline
 &&&\\
\sharp 3  & N_3=1& (F^3_{x_1 x_2},F^3_{y_1 y_2})=(1,1)& 
 [x_3,y_3] \\
               &&&\\
\hline 
\hline
&&&\\
\sharp 4  & N_4=1& (F^4_{x_2 x_3},F^4_{y_2 y_3})=(1,1)& 
 [x_1,y_1] \\
&&&\\
\hline
&&&\\
\sharp 5 & N_5=1& (F^5_{x_1 x_3},F^5_{y_1 y_3})=(1,1) & 
 [x_2,y_2]\\
&&&\\
\hline
&&&\\
\sharp 6  & N_6=1& (F^6_{x_2 y_3},F^6_{x_3 y_2})=(1,1) & 
 [x_1,y_1]\\
&&&\\
\hline
\end{array}
$$
\end{center}
This table presents a configuration of oblique fluxes that lead to the stabilization of all complex structure and all off-diagonal K\"ahler moduli of the torus.  The last column shows the localization of the 5-brane tadpoles induced by the flux quanta. The resulting shape is a square torus $T^2 \times T^2 \times T^2$ with $\tau_{ij} = i\delta_{ij}$.

 \bibliographystyle{fullsort}
 \providecommand{\href}[2]{#2}\begingroup\raggedright\endgroup


\begin{thebibliography}{10}

\bibitem{Douglas:2006es}
M.~R. Douglas and S.~Kachru, ``Flux compactification,''
\href{http://www.arXiv.org/abs/hep-th/0610102}{{\tt hep-th/0610102}} and references therein

\bibitem{Review}
R.~Blumenhagen, B.~Kors, D.~Lust, and S.~Stieberger, ``Four-dimensional string
  compactifications with d-branes, orientifolds and fluxes,''
\href{http://www.arXiv.org/abs/hep-th/0610327}{{\tt hep-th/0610327}} 

\bibitem{openAS}
C.~Angelantonj and A.~Sagnotti, ``Open strings,'' {\em Phys. Rept.} {\bf 371}
  (2002) 1--150,
\href{http://www.arXiv.org/abs/hep-th/0204089}{{\tt hep-th/0204089}}.

\bibitem{Will:2001mx}
C.~M. Will, ``The confrontation between general relativity and experiment,''
  {\em Living Rev. Rel.} {\bf 4} (2001) 4,
\href{http://www.arXiv.org/abs/gr-qc/0103036}{{\tt gr-qc/0103036}}.

\bibitem{Gomis:2005wc}
J.~Gomis, F.~Marchesano, and D.~Mateos, ``An open string landscape,'' {\em
  JHEP} {\bf 11} (2005) 021,
\href{http://www.arXiv.org/abs/hep-th/0506179}{{\tt hep-th/0506179}}.

\bibitem{Lust:2005bd}
D.~Lust, P.~Mayr, S.~Reffert, and S.~Stieberger, ``F-theory flux,
  destabilization of orientifolds and soft terms on d7-branes,'' {\em Nucl.
  Phys.} {\bf B732} (2006) 243--290,
\href{http://www.arXiv.org/abs/hep-th/0501139}{{\tt hep-th/0501139}}.

\bibitem{Angelantonj:2003zx}
C.~Angelantonj, R.~D'Auria, S.~Ferrara, and M.~Trigiante, ``K3 x t**2/z(2)
  orientifolds with fluxes, open string moduli and critical points,'' {\em
  Phys. Lett.} {\bf B583} (2004) 331--337,
\href{http://www.arXiv.org/abs/hep-th/0312019}{{\tt hep-th/0312019}}.

\bibitem{Angelantonj:2000hi}
C.~Angelantonj, I.~Antoniadis, E.~Dudas, and A.~Sagnotti, ``Type-i strings on
  magnetised orbifolds and brane transmutation,'' {\em Phys. Lett.} {\bf B489}
  (2000) 223--232,
\href{http://www.arXiv.org/abs/hep-th/0007090}{{\tt hep-th/0007090}}.

\bibitem{Blumenhagen:2000wh}
R.~Blumenhagen, L.~Goerlich, B.~Kors, and D.~Lust, ``Noncommutative
  compactifications of type i strings on tori with magnetic background flux,''
  {\em JHEP} {\bf 10} (2000) 006,
\href{http://www.arXiv.org/abs/hep-th/0007024}{{\tt hep-th/0007024}}.

\bibitem{Cremades:2007ig}
D.~Cremades, M.~P. Garcia~del Moral, F.~Quevedo, and K.~Suruliz, ``Moduli
  stabilisation and de sitter string vacua from magnetised d7 branes,''
\href{http://www.arXiv.org/abs/hep-th/0701154}{{\tt hep-th/0701154}}.

\bibitem{Cremades:2004wa}
D.~Cremades, L.~E. Ibanez, and F.~Marchesano, ``Computing yukawa couplings from
  magnetized extra dimensions,'' {\em JHEP} {\bf 05} (2004) 079,
\href{http://www.arXiv.org/abs/hep-th/0404229}{{\tt hep-th/0404229}}.

\bibitem{Dine:1987xk}
M.~Dine, N.~Seiberg, and E.~Witten, ``Fayet-iliopoulos terms in string
  theory,'' {\em Nucl. Phys.} {\bf B289} (1987)
589.

\bibitem{Atick:1987gy}
J.~J. Atick, L.~J. Dixon, and A.~Sen, ``String calculation of fayet-iliopoulos
  d terms in arbitrary supersymmetric compactifications,'' {\em Nucl. Phys.}
  {\bf B292} (1987)
109--149.

\bibitem{BLT}
R.~Blumenhagen, D.~Lust, and T.~R. Taylor, ``Moduli stabilization in chiral
  type iib orientifold models with fluxes,'' {\em Nucl. Phys.} {\bf B663}
  (2003) 319--342,
\href{http://www.arXiv.org/abs/hep-th/0303016}{{\tt hep-th/0303016}}.

\bibitem{Cvetic:2001nr}
M.~Cvetic, G.~Shiu, and A.~M. Uranga, ``Chiral four-dimensional n = 1
  supersymmetric type iia orientifolds from intersecting d6-branes,'' {\em
  Nucl. Phys.} {\bf B615} (2001) 3--32,
\href{http://www.arXiv.org/abs/hep-th/0107166}{{\tt hep-th/0107166}}.

\bibitem{AM}
I.~Antoniadis and T.~Maillard, ``Moduli stabilization from magnetic fluxes in
  type i string theory,'' {\em Nucl. Phys.} {\bf B716} (2005) 3--32,
\href{http://www.arXiv.org/abs/hep-th/0412008}{{\tt hep-th/0412008}}.

\bibitem{AKM2}
I.~Antoniadis, A.~Kumar, and T.~Maillard, ``Magnetic fluxes and moduli
  stabilization,''
\href{http://www.arXiv.org/abs/hep-th/0610246}{{\tt hep-th/0610246}}.

\bibitem{GarciadelMoral:2005js}
M.~P. Garcia~del Moral, ``A new mechanism of kahler moduli stabilization in
  type iib theory,'' {\em JHEP} {\bf 04} (2006) 022,
\href{http://www.arXiv.org/abs/hep-th/0506116}{{\tt hep-th/0506116}}.

\bibitem{Freed:1999vc}
D.~S. Freed and E.~Witten, ``Anomalies in string theory with d-branes,''
\href{http://www.arXiv.org/abs/hep-th/9907189}{{\tt hep-th/9907189}}.

\bibitem{ADM}
I.~Antoniadis, J.~P. Derendinger, and T.~Maillard, to appear. 
\\
I. Antoniadis, Talk at String Phenomenology Conference, Rome Frascati, June 2007 and 4th regional meeting  in string theory, Patras, June 2007. 

\bibitem{Strominger:1985it}
A.~Strominger and E.~Witten, ``New manifolds for superstring
  compactification,'' {\em Commun. Math. Phys.} {\bf 101} (1985)
341.

\bibitem{Bertolini:2005qh}
M.~Bertolini, M.~Billo, A.~Lerda, J.~F. Morales, and R.~Russo, ``Brane world
  effective actions for d-branes with fluxes,'' {\em Nucl. Phys.} {\bf B743}
  (2006) 1--40,
\href{http://www.arXiv.org/abs/hep-th/0512067}{{\tt hep-th/0512067}}.

\bibitem{DiVecchia:1999rh}
P.~Di~Vecchia and A.~Liccardo, ``D branes in string theory. i,'' {\em NATO Adv.
  Study Inst. Ser. C. Math. Phys. Sci.} {\bf 556} (2000) 1--59,
\href{http://www.arXiv.org/abs/hep-th/9912161}{{\tt hep-th/9912161}}.

\bibitem{Bianchi:2005yz}
M.~Bianchi and E.~Trevigne, ``The open story of the magnetic fluxes,'' {\em
  JHEP} {\bf 08} (2005) 034,
\href{http://www.arXiv.org/abs/hep-th/0502147}{{\tt hep-th/0502147}}.

\bibitem{Font:2006na}
A.~Font, L.~E. Ibanez, and F.~Marchesano, ``Coisotropic d8-branes and
  model-building,'' {\em JHEP} {\bf 09} (2006) 080,
\href{http://www.arXiv.org/abs/hep-th/0607219}{{\tt hep-th/0607219}}.

\bibitem{Anastasopoulos:2006hn}
P.~Anastasopoulos, M.~Bianchi, G.~Sarkissian, and Y.~S. Stanev, ``On gauge
  couplings and thresholds in type i gepner models and otherwise,''
\href{http://www.arXiv.org/abs/hep-th/0612234}{{\tt hep-th/0612234}}.

\bibitem{Marino:1999af}
M.~Marino, R.~Minasian, G.~W. Moore, and A.~Strominger, ``Nonlinear instantons
  from supersymmetric p-branes,'' {\em JHEP} {\bf 01} (2000) 005,
\href{http://www.arXiv.org/abs/hep-th/9911206}{{\tt hep-th/9911206}}.

\bibitem{Bianchi:2005sa}
M.~Bianchi and E.~Trevigne, ``Gauge thresholds in the presence of oblique
  magnetic fluxes,'' {\em JHEP} {\bf 01} (2006) 092,
\href{http://www.arXiv.org/abs/hep-th/0506080}{{\tt hep-th/0506080}}.

\bibitem{Blumenhagen:2000dc}
R.~Blumenhagen and A.~Font, ``Dilaton tadpoles, warped geometries and large
  extra dimensions for non-supersymmetric strings,'' {\em Nucl. Phys.} {\bf
  B599} (2001) 241--254,
\href{http://www.arXiv.org/abs/hep-th/0011269}{{\tt hep-th/0011269}}.

\bibitem{Dudas:2000ff}
E.~Dudas and J.~Mourad, ``Brane solutions in strings with broken supersymmetry
  and dilaton tadpoles,'' {\em Phys. Lett.} {\bf B486} (2000) 172--178,
\href{http://www.arXiv.org/abs/hep-th/0004165}{{\tt hep-th/0004165}}.

\bibitem{Angelantonj:2007ts}
C.~Angelantonj and E.~Dudas, ``Metastable string vacua,''
\href{http://www.arXiv.org/abs/arXiv:0704.2553 [hep-th]}{{\tt arXiv:0704.2553
  [hep-th]}}.

\end{thebibliography}

\end{document}